\documentclass[preprint,12pt]{article}

\usepackage[font=small]{caption}
\usepackage{amssymb, amsmath, amsbsy} 
\usepackage{slashed}
\usepackage{graphicx}
\usepackage{float}%%%% imagen acá te quiero
\usepackage{dsfont}%%%%% letras doble linea %%%%% 
\usepackage{color}
\usepackage[affil-it]{authblk}
\usepackage{authblk}%%%% more than one author
\usepackage{rotating}%%% para rotar figuras
\usepackage{booktabs}%%%% espacio entre renglones de tablas
\usepackage[left=2cm,right=2cm,top=3cm,bottom=3cm]{geometry}
\usepackage{xcolor}

\author[1,2]{J. Lorenzo D\'iaz-Cruz\thanks{\texttt{jldiaz@fcfm.buap.mx}}}

\title{{\bf Solving the Naturalness Problem with Feeble Coupled Sectors}}

\affil[1]{CIIEC and CIFFU, BUAP} 
\affil[2]{ Facultad de Ciencias F\'isico - Matem\'aticas, BUAP \protect\\Apdo. Postal 1364, C.P. 72000, Puebla, Pue. M\'exico}

\begin{document} 

\maketitle

\begin{abstract}
The discovery of a light Higgs boson means that whatever form new physics takes, it should keep stable the Higgs mass. Besides the well-known solutions to the naturalness problem (Supersymmetry, Conformal symmetry, Compositeness, etc),  models that include heavy  particles with feeble couplings to the Standard Model (SM) can be considered natural, since the corrections to the Higgs mass remains of the order of the electroweak (EW) scale.
This solution can be used for model building too, with realizations that include the see-saw  mechanism for neutrino masses and FIMP dark matter models, but it also holds for generic sectors that have Planck-suppressed couplings with the SM. One can also incorporate this solution within the SMEFT framework; the corresponding higher-dimensional operators induce small corrections to both the Higgs mass and its self-coupling, a prediction that can be tested at a future Higgs factory. We present a natural extension of the SMEFT that describes corrections to the SM, while also including a Feeble Coupled Sector aimed to account for the dark cosmos, with predictions for new signals that can be tested too.

\end{abstract}

\newpage

{\bf{1. Introduction.}} Finding a solution to the hierarchy or naturalness problem, and exploring the corresponding models, have played an important role for the development of particle physics. The identification of quadratic divergences as a potential peril  for elementary scalars \cite{Susskind:1978ms}, motivated the search for new theories where such problem could be eliminated. The importance of those solutions, such as supersymmetry, composite and technicolor models, should not be underestimated, both for conceptual and experimental aspects.   In the first place they provided roads to look for and explore 
models that addressed such problem; furthermore those models, despite its failure or success,  justified the experimental search for the Higgs boson, which many doubted it actually existed after the Standard Model (SM) was proposed. The development of radiative corrections helped to corner the Higgs in a mass region at the reach of collider experiments, thus from LEP we learned that the Higgs mass has to be above 105 GeV, while Tevatron helped to exclude the range around the ZZ threshold. Finally, the debate about whether nature likes or not elementary scalars, was closed after  the discovery of the Higgs boson at the CERN LHC, with a mass $m_h=125$ GeV and properties in agreement with the SM.

 Further conceptual developments helped to understand better the nature of quadratic divergences. In particular, we learned that the conformal symmetry \cite{Bardeen:1995kv} could help to understand the origin of the Higgs mass as a soft-breaking effect,
 and  therefore the SM itself can be considered as natural. Furthermore,  after having provided a method to identify quadratic divergences within the dimensional regularization method \cite{Veltman:1980mj}, Veltman departed from the common lore and claimed that there are no quadratic divergences in the SM \cite{Veltman:1994aq}.  Thus,  it is appropriate to say that the SM is natural, as the radiative contribution to the Higgs mass coming from its heavier particle, the top quark, keeps  the Higgs mass of the order of the EW scale, suggesting the absence of the hierarchy
 problem within the SM \cite{Wells:2013tta,Jegerlehner:2018zxm}. 
 
 The real problem appears  for extensions of the SM that include heavy states, with mass and couplings given by 
 $M_X$ and $g_x$, respectively; after integrating them out, they leave a potential large correction to the Higgs mass ($m_h$), which is of the order $\delta m^2_h \simeq g_x M^2_X$, and for $M_X >> m_W$, there appears to be a large correction to the Higgs mass. The traditional solutions to this problem assume that the coupling constant $g_X$ of the new heavy particles has a value similar to the other couplings in the SM.  In SUSY models, for instance, one has both fermions and sfermions  \cite{Haber:1985fe}, and their contribution to the Higgs mass have opposite signs, and thus cancel each other. In the conformal solution \cite{Meissner:2006zh}, the tree-level contribution vanishes, and the Higgs mass itself is generated as a radiative effect. 
    But we also know that, apart from the naturalness problem, there are other problems in the SM (flavor, generations, unification), while the evidence for  neutrino masses  and the dark cosmos, i.e. dark matter and dark energy, as well as the early universe inflation and the matter-antimatter asymmetry, which can not be explained by the SM. Thus, it seems there are good reasons to expect that some form of new physics should exist.    One could go ahead studying models that address some Beyond the SM issues, without caring about the solution to the naturalness problem, hoping that some additional new physics will bring a solution to the problem.  Alternatively, we can take to the heart the issue of naturalness, and consider only extended models that incorporate a solution to the problem, as it was done in the past with supersymmetry or other BSM scenarios. 

So far we have a few experimental facts that could guide us in the search for BSM physics, namely, the detection of a light Higgs, the absence of  BSM signals at LHC and the lack of direct evidence for dark matter WIMPS.    In this paper we shall try to incorporate these three hints into a single framework. We shall assume that there exists heavy particles beyond the SM, which play a role for the solution of some of its open problems, that induce a correction to the Higgs mass of the order $\delta m^2_h \simeq g_x M^2_X$, but this is natural because rather than having a coupling of O(1), the coupling $g_x$ is feeble, such that $g_x M^2_X \simeq m^2_h$, which can be considered as a solution to the naturalness problem. We argue that this solution with q natural Feeble Coupled Sector (FECOS),  should be consider on the same foot, or perhaps even more solid, than the other ones, because it is based on the current observations; furthermore, it provides a framework  for model building, like SUSY or Techinicolor.  FECOS  solution can occur in many models, such as the Type-I see-saw model for neutrino masses, or in models where a Feeble-interacting massive particle (FIMP) is the candidate for dark matter; moreover it can also work for generic models that include heavy states with suppressed couplings to the SM, including models with the gravity portal.

  We can consider a generic FECOS model, and notice that when its heavy modes are integrated out, it should leave an effective Lagrangians extension of the SM, which includes a whole tower of higher dimensional operators suppressed by a scale 
  $\Lambda\simeq M_X $. Within this affective Lagrangian, there should be no problem with quadratic divergences, and the corrections to the Higgs mass should be at most of the order $O(v/\Lambda)$, and similarly for the Higgs self coupling, a prediction that can be tested at a future Higgs factory. Furthermore, one can build a specific model  that include a FECOS aimed to describe aspects of the Dark Cosmos (DC), for instance by including extra scalars and fermions to account for dark matter and the inflaton. This model, which we call  the DC-SMEFT, includes  renormalizable Lagrangians for both SM and DC sectors, but it also contains higher-dimensional operators for both sectors, including their mixture. This model provides some distinctive signatures that could be probed in diverse fronts, ranging form colliders to cosmology.

{\bf{2. Natural models with Feeble Coupled Sector.}}  
 To discuss the problem of naturalness, let us consider  a theory that includes a SM-like Higgs boson ($h$) which interacts with a heavy scalar $S$, gauge field $V$ and fermion $F$. The corresponding expression for the associated quadratic divergences is:  
  \begin{equation}
 m^2_h(\Lambda, \mu)= m^2_h(\Lambda) + \sum_{X=S,V,F} (-1)^{2J_x} (2J_x+1) \frac{g_x}{16\pi^2}
 [\Lambda^2 - m^2_h(\Lambda) \log \frac{\Lambda^2}{\mu^2}]
 \end{equation}
where $\mu$ denotes the renormalization scale, and  $\Lambda$ denotes the momentum cutoff.
Nowadays we understand better the meaning of this equation. Namely, when $\Lambda$ corresponds to the UV cutoff, one simply has to renormalize this effect on the Higgs mass. Given that the  SM is a renormalizable theory, even when one takes $\Lambda \to \infty$, no observable effect is left. However, the real problem resides when $\Lambda$ represents the mass of a heavy particle  i.e. $\Lambda=M$, and when it is integrated out, it leaves a correction to the Higgs proportional to the scale $\Lambda=M$, which could be very large. However, having a large but finite value for $\Lambda$, is different from having to deal with a divergent term, even when it is of the order of the Planck mass, i.e. $\Lambda < M_{pl}$; this was used in the past to constrain on the Higgs boson mass \cite{Kolda:2000wi}. In order to cancell the correction to the Higgs mass, which has to be the order of the EW scale, one can have some relationship among the couplings or the masses \cite{Kobakhidze:2014afa}. The traditional solutions to this problem assume that the coupling constant $g_X$ has a value similar to the other couplings in the SM, and imply the existence of new heavy particles, which can be searched  at collider or cosmolgy. SUSY \cite{Haber:1985fe} corresponds to the first case, and conformal symmetry is of the second type, and these are two well-known possibilities to cancel the quadratic divergence. 

Here we want to argue that there is also another possibility to bring under control the naturalness problem. 
We shall assume that some generic extensions of the SM includes heavy particles, which induce a natural correction to the Higgs mass, of the order $O(m_W)$, because rather than having a coupling $g_x$ of O(1), we assume  that this coupling is feeble, such that $g_x M^2_X \simeq m^2_h$ even for $M_X >> m_W$, which solves the naturalness problem. We argue that this resolution to the naturalness problem with Feeble-Coupled Sectors (FECOS),  should be consider on the same foot as the other solutions, or perhaps even better, because its motivation is based on the current observations.
One example that proves this mechanism is feasible corresponds to the SM augmented with neutrino masses obtained through the see-saw mechanism \cite{Vissani:1997ys}.  In this case it happens that the correction to the Higgs mass is given by $\delta m^2_h= y_\nu M^2/16 \pi^2$,
with the neutrino Yukawa coupling being given by $y_\nu = M m_\nu/v$, for $m_\nu=0.05$ GeV, $\delta m^2_h$ is of the order of the EW scale even for $M \simeq 10^8$; and the $\nu SM$ is natural even when it contains a heavy RH neutrino.
Another class of models that falls within the FECOS category is the so called Feeble Interacting massive particle (FIMP), which
is a viable solution to the dark matter problem \cite{Hall:2009bx,McDonald:2001vt}. The viability of this scenario is based on a different calculation of the DM relic density, the freeze-in mechanism, which relies on the feeble coupling of the FIMP candidate.

Furthermore,  the FECOS solution to the naturalness problem, can be a generic one, it is of a bottom-up type, which accommodates a light Higgs boson, and within specific models it can explain neutrino masses and dark matter. In the FECOS models we can have
heavy particles, but with very small couplings, even suppressed by the Planck mass,  $m^2/M_{pl}$, as it occurs in supergravity. We can also consider models with a gravitational sector that includes other FECOS fields that couple gravitationally with the  SM, like the heavy graviton that appears in extra dimensional models  \cite{Barman:2022qgt}, with couplings suppressed by the volume of the extra-dimensional space, or by the curvature of the extra-dimension in Randall-Sundrum 5D models, where the SM resides on a 
EW brane, while the graviton and other fields propagate in the bulk. It is also interesting to discuss the effect of gravity on quadratic divergences, but as shown in \cite{Meissner:2018mvq}, these corrections do not affect the Higgs mass.
\bigskip

 {\bf{3. Naturalness in the SMEFT and the Dark Cosmos.}}  We would like to apply the FECOS solution of naturalness, 
 to discuss some aspects of the Dark Cosmos (DC), in particular dark matter. Our complete model will contain the SM sector, as well as a new DC sector of the FECOS type.  One can look  directly at the effects of the new sector, for instance in neutrino physics or dark matter searches. It is also possible to calculate their effects on the properties of the SM particles, in particular the Higgs boson, which can be described with an effective Lagrangian.  In fact, we should  notice that the level of divergence in the fundamental theory could be less severe than one can expect  from the effective lagrangian \cite{Biswas:2020abl}.
 The effective Lagrangian  includes renormalizable parts, for both the SM and for the new particles of the FECOS type, but it 
 contains as  well higher-dimensional operators for both sectors.  For instance, the renormalizable part can include a mixing term of the form  $(\Phi^\dagger \Phi) S^2$,  between the SM Higgs doublet $\Phi$ and an extra singlet $S$. Then, after integrating out the singlet $S$, one gets a new operator of the form $(\Phi^\dagger \Phi)^3$, which modifies the Higgs properties.  However this operator comes from a sector with  suppressed coupling, and will be highly suppressed. Thus, within the FECO models the modification of the Higgs properties, such as the 3-particle self-coupling, will only show tiny deviations from the SM, probably unobservable. However,  besides the naturalness problem, there are other open issues in the SM, that also demand some physics beyond the SM, such as the origin of flavor, CPV, unification, etc.  As we do not have a solution to these problems, for the sake of generality, we shall consider  an effective Lagrangian that describes the  effects of those new particles and interactions with the SM particles. This model could include extra fermions $\psi_a$  and some scalars $S_i$,  within a minimal setting of the FECOS type. 
 So far we do not have enough information to include gauge  interactions for these particles, unlike the SM, so we can start by including a number of fields that play a role for the resolution of the problems of Dark matter, inflation, and possibly the matter-antimatter asymmetry, similar to the $\nu SM$ \cite{Davoudiasl:2004be}.
We call this model the Dark Cosmos extension of the SMEFT (DC-SMEFT), and its lagrangian has the general form:
  
 \begin{equation}
{\cal{L}}_{DC-SMEFT}= {\cal{L}}_{SM} + {\cal{L}}_{DC} +  \sum_{i,d} 
[ \frac{\alpha_{i,d}}{\Lambda^{d-4}}  O^{sm}_{d,i} + \frac{\beta_{i,d}}{\Lambda^{d-4}}  O^{sm}_{d,i}]
 \end{equation}
 
 Here  ${\cal{L}}_{SM}$ and  ${\cal{L}}_{DC}$ represent the renormalizable Lagrangians for the SM and Dark cosmos; for which we can include kinetic terms, Yukawa interactions and an scalar potential in ${\cal{L}}_{DC}$.
 The higher d-dimensional operators $O^{SM}_{i,d}$, involve only the SM fields, while  $O^{DC}_{i,d}$ involve the extra
 fields, as well as its mixing with SM sector. For $O^{SM}_{i,6}$ we can include the whole list of dimension-six operators 
 \cite{Grzadkowski:2010es}, and as mentioned before, we are interested in the operators that modify the Higgs properties,
 which is discussed for the complete set of associated operators in \cite{Asteriadis:2022ras}. One possible realization of this approach is the specific FIMP model presented in \cite{Belanger:2021slj}, where the DC sector includes three right-handed neutrinos ($N_i$) and  a real scalar ($S$), two of the RH neutrinos participate in the see-saw mechanism to generate light neutrino masses, while the third one is the DM candidate. The vev of the scalar singlet generates the light neutrino masses through dimension-5 operators, and several scenarios with a stable FIMP are found to satisfy current constraints. 
  
 The associated phenomenology of these models can be studied, for instance by looking for small deviations from the SM predictions. However, it could be even more interesting to look for rare effects that are included in the higher-dimensional operators. For instance, one can look at the effects of the operators included in the model of  \cite{Belanger:2021slj}, but for an unstable FIMP particle, which can provide a distinctive signature of the model. In particular, one has a dimension-five term of the form: $L\tilde{\Phi}N_i S$, and after both the SM Higgs doublets $\Phi$ and the singlet $S$, aquire a vev, there is a mixing between the third RH neutrino,  which is assumed to be the FIMP DM, and this will induce decays of the dark matter. The corresponding life-time should be
larger than the age of the universe, for this scenario to be a viable, but even after such constraint is satisfied, it is possible to
observe effects of decaying DM. One could also decouple the neutrino mass problem from the Dark Matter issue, by adding an extra fermion, different from the RH neutrinos, which will play de role of FIMP dark matter, while the singlet scalar could still participate in the generation of the see-saw mechanism.   Detailed analysis of the model is under study. A different model, that also shows the viability of the scenario with a decaying FIMP is presented in \cite{Covi:2020pch}.

\bigskip

 {\bf{4. Conclusions.}} 
We presented a discussion of the naturalness problem, and argued that  besides the well-known solutions to the naturalness problem (such as supersymmetry, Conformal symmetry, Compositeness, etc),  there is also another solution,  namely the case when a new sector with heavy particles have feeble-couplings with the Standard Model (SM), which induce natural corrections to the Higgs mass.  This solution with Feeble-coupled sectors (FECOS), can be used for model building too, for instance in the see-saw  mechanism for neutrino masses, or the FIMP dark matter model, but it is a generic one, and one can include models with heavy states that have Planck-suppressed couplings with the SM. 
The FECOS solution to the naturalness problem, can be a generic one, it is of a bottom-up type, which accommodates a light Higgs boson, and within specific models it can explain neutrino masses and dark matter, while being in agreement  with the absence of signals in collider and direct search for DM.  

Further aspects of naturalness can be discussed within the SMEFT framework, with higher-dimensional operators that induce small corrections to both the Higgs mass and its self-coupling, and measuring it provides a strong motivation for a future Higgs factory. We present a natural extension of the SMEFT that includes a FIMP dark matter candidate, which can provide new signals that test this whole new framework. In particular, decay of the DM particle could be a distinctive signature of this model.

\end{document}